\documentclass[doublecol]{epl2}

\usepackage{amsmath}

\title{Fermi surface arcs and the infrared conductivity of
underdoped YBa$_{2}$Cu$_{3}$O$_{6.50}$}

\author{J. Hwang\inst{1} \and J. P. Carbotte\inst{1,2} \and T. Timusk\inst{1,2}}

\institute{
  \inst{1} Department of Physics and Astronomy, McMaster
University, Hamilton, Ontario L8S 4M1, Canada\\
  \inst{2} The
Canadian Institute for Advanced Research, Toronto, Ontario M5G
1Z8, Canada} 

\abstract{We reanalyze the data on the in-plane far infrared
conductivity of underdoped orthoII YBa$_{2}$Cu$_{3}$O$_{6.50}$
(YBCO$_{6.50}$) in terms of a model in which a pseudogap opens on
part of the Fermi surface with the remaining ungaped piece
proportional to the temperature. The motivation for our model
comes from recent angle-resolved photoemission spectroscopy data
in underdoped Bi$_2$Sr$_2$CaCu$_2$O$_{8+\delta}$ (Bi-2212), which
have revealed the existence of arcs on the Fermi surface. We find
the optical data to be consistent with arc formation. In addition
we find some evidence that the electronic states lost below the
pseudogap energy $\Delta_{pg}$ are recovered in the energy region
immediately above it at least for temperatures near $T_c$.}

\begin{document}

\maketitle

In going from the overdoped side of the cuprate phase diagram to
the underdoped side, a new phenomenon known generally as the
formation of a pseudogap is seen\cite{timusk99}. The pseudogap
manifests itself in many properties and generally has the
signature of a decrease in the electronic density of state (DOS)
around the Fermi energy. There is no consensus as to its exact
microscopic origin. Some observations are consistent with a
preformed pair model\cite{emery95} with the pseudogap associated
with the binding energy of the pairs at the pseudogap temperature
$T^{*}$ and the superconducting temperature $T_c$ the point at
which phases lock in and long range order results. Another
influential model is a possible competing order such as
d-density\cite{chakravarty01,dora02,benfatto06,aristov05,kim02,kim02a}
wave which set in at $T^{*}$ and coexists with the superconducting
order parameter below $T_c$. In this model it may be reasonable to
expect two distinct gap scales and some recent
data\cite{krasnov00,letacon06} have pointed to this possibility.

In the absence of consensus on microscopic origin,
phenomenological models have played an important role in
understanding and correlating data. An example is the specific
heat data\cite{loram98,loram01}. The observed trend from overdoped
to underdoped can be understood \cite{loram98,loram01} over
several families of cuprates by simply introducing a gap
$\Delta_{pg}$ in the DOS which obeys a mean field temperature
dependence. More recently, analysis of angle-resolve photoemission
spectroscopy (ARPES) data have given a more detailed
picture\cite{kanigal06} of pseudogap formation on the Fermi
surface as a function of temperature. A finite pseudogap first
opens at $T^{*}$ but forms only in a region around the antinodal
direction leaving an ungaped Fermi surface arc around the nodal
direction of length proportional to the reduced temperature $t =
T/T^{*}$. Here we confine our attention to the normal state region
(above $T_c$) and study how the formation of Fermi surface arcs
modifies optical properties.

Optical conductivity measurements have given a wealth of
information on electron dynamics in the cuprates. While early on
the pseudogap was seen clearly as a real gap in the real part of
the c-axis\cite{homes93,homes95} optical conductivity, its
signature in the ab-plane data was less clear and largely confined
to a general observation that the optical scattering rate seem to
show an additional decrease with decreasing temperature at low
frequency as compared with overdoped samples\cite{puchkov96}. Only
recently has the situation been further clarified when the real
part of the optical self-energy was analyzed in more
detail\cite{hwang07}. Underdoped samples show a characteristic
evolution with decreasing temperature quite distinct from the
overdoped case\cite{hwang04,hwang07a,hwang06}. A hat like peak
structure develops at an energy of the order of the pseudogap
energy, clearly seen above a smooth structureless background. It
was recognized that this easily identified\cite{hwang07} feature
could be understood in a simple DOS model with a gap below
$\Delta_{pg}$ and the missing states piled up in the energy region
just above the pseudogap which we will refer to here as the
recovery region. These modification of the DOS show up very
directly in the real and imaginary part of the optical self-energy
$\Sigma^{op}(\omega)$\cite{mitrovic85,sharapov05,knigavko05,knigavko06}.

Consider coupling to a single Einstein mode of energy $\omega_E$.
Without a pseudogap (constant electronic DOS) the optical
scattering rate $1/\tau^{op}(\omega)$ will be zero till $\omega =
\omega_E$ after which it rises from zero according to the law
$(\omega-\omega_E)/\omega$\cite{carbotte05}. With a pseudogap it
will be zero till $\omega = \omega_E + \Delta_{pg}$ and rise after
this according to
$(\omega-\omega_E-\Delta_{pg})/\omega$\cite{hwang06}. On the other
hand if a recovery region is included above $\omega = \Delta_{pg}$
to conserve electronic states, there will be additional scattering
in the recovery region due to the extra available states. This
will make $1/\tau^{op}(\omega)$ rise more rapidly than it would
have without the extra states. This rapid rise in
$1/\tau^{op}(\omega)$ translates into the formation of a
logarithmic like structure in Kramers-Kronig transform. Monitoring
the growth of this structure in the real part of the optical
self-energy $\Sigma^{op}_{1}(\omega)$ gives us a means to trace
pseudogap formation as does the steepness of $1/\tau^{op}(\omega)$
above the energy of the gap plus boson.

There have been many studies of the boson spectra of the cuprates
as revealed by tunneling\cite{zasadzinski06}, ARPES\cite{valla06}
and
optics\cite{hwang06,carbotte99,schachinger00,schachinger06,dordevic05,hwang07b,hwang07c}
in which a spectral density function $I^2\chi(\omega)$ is
recovered, where $I$ is a coupling constant between electrons and
a boson and $\chi(\omega)$ is the bosonic spectral density. For
optimally and overdoped cases, coupling to the sharp resonance at
an energy corresponding to the spin-1 collective mode measured in
neutron scattering is seen. This resonance starts to exist above
$T_c$ in the underdoped case and its intensity (area under the
peak) increases with decreasing temperature. For the specific case
of orthoII YBCO$_{6.50}$ this has been studied by Hwang {\it et
al.}\cite{hwang06}.

%
%
\begin{figure}[t]
  \vspace*{-0.5 cm}%
  \centerline{\includegraphics[width=3.5 in]{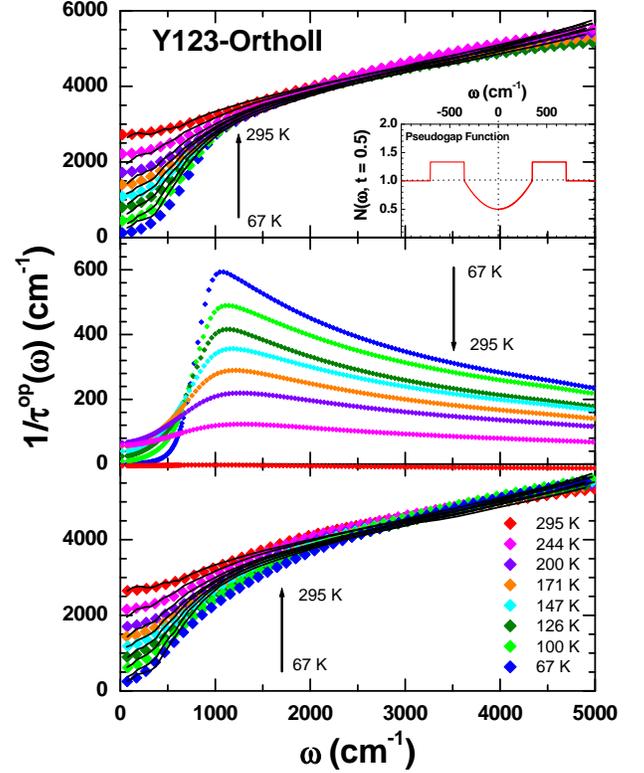}}%
  \vspace*{-0.5 cm}%
\caption{(color online) Top frame, optical scattering rate
$1/\tau^{op}(\omega)$ in cm$^{-1}$ as a function of $\omega$ also
in cm$^{-1}$ for the eight temperatures shown (normal state). The
heavy solid squares are fits to the data (solid lines). The inset
shows the frequency dependent of $\tilde{N}(\omega,t)$ for $t=$
0.5, with quadratic frequency dependence below $\Delta_{pg}$ and a
constant above it for $\omega \in (\Delta_{pg},2\Delta_{pg})$.
Electronic states are conserved. Bottom frame, same as top but
previous fit to the data obtained in reference [20] without
including a recovery region above to pseudogap energy which is
chosen to conserve state. Middle frame, difference between top
frame and scattering rates obtained when recovery region is
excluded, keeping other parameter the same (as top frame fits).}
 \label{Fig1}
\end{figure}
In this paper we reanalyze the data on orthoII YBCO$_{6.50}$
allowing for the opening of a pseudogap which covers more of the
Fermi surface, as the temperature is lowered below $T^{*}$. We ask
whether or not the data is consistent with a linear decrease in
the length of the remaining arcs as a function of the reduced
temperature $t = T/T^{*}$ as it is seen in the ARPES data in
underdoped Bi-2212\cite{kanigal06}. Our analysis of the optical
data proceeds through the optical scattering rate
$1/\tau^{op}(T,\omega)$. For a system with an energy dependent
self consistent electronic density of state $N(\omega)$ which is
related to the imaginary part of the Green's function
$G(\underline{k},\omega)$ by
$N(\omega)=1/\pi\sum_{\underline{k}}[-ImG(\underline{k},\omega)]$
the optical scattering rate can be written
approximately\cite{sharapov05} as
\begin{equation}
\begin{split}
1/\tau^{op}(T,
\omega)=&\frac{2\pi}{\omega}\int^{\infty}_{0}d\Omega
I^2\chi(\Omega) \int^{\infty}_{-\infty}d\epsilon
\tilde{N}(\epsilon-\Omega)\times \\
&[n(\Omega)+f(\Omega-\epsilon)][f(\epsilon-\omega)-f(\epsilon+\omega)]
\end{split} \label{eq1}
\end{equation}
Here $n(\Omega)=1/[e^{\beta \Omega}-1]$ is the Bose-Einstein
thermal distribution, $f(\epsilon)=1/[e^{\beta (\mu-\epsilon)}+1]$
the Fermi-Dirac with $\beta$ the inverse temperature and $\mu$ the
chemical potential, and $I^2\chi(\omega)$ the electron-boson
spectral density. In Eq. \ref{eq1} $\tilde{N}(\epsilon)$ is the
symmetrized electronic density of states
$[N(\epsilon)+N(-\epsilon)]/2$. Eq. \ref{eq1} reduces properly to
that of Shulga {\it et al.}\cite{shulga91} in the limit of a
constant density of states. It further corresponds to the formula
of Allen\cite{allen71} at zero temperature. The data for
$1/\tau^{op}(T, \omega)$ at eight temperatures in the normal state
of orthoII YBCO$_{6.50}$ are reproduced\cite{hwang06} in Fig.
\ref{Fig1} (top and bottom frames) as (dark) solid lines. In all
cases the data is fit (solid diamonds of different colors as
indicated in the figure) with an electron-boson spectral density
$I^2\chi(\omega)$ consisting of a broad Millis-Monien-Pines (MMP)
background\cite{millis90} of the form $I_0\:
\omega/(\omega^2+\omega_{sf}^2)$, where $I_0$ is a coupling
between spin fluctuation and charge carriers and $\omega_{sf}$ is
a typical spin fluctuation energy. In addition a sharp resonance
peak modelled by $I_p/[\sqrt{2\pi}(d/2.35)]
e^{-(\omega-\omega_{p})^2/[2(d/2.35)^2]}$ (where $I_p$ is the area
under the peak, $\omega_p$ is its center frequency, and $d$ is its
width) is included and its position in energy fixed at 31 meV
chosen to agree with the energy of the measured spin-1 resonance
of inelastic neutron scattering\cite{stock05}. In addition a
pseudogap is included with $\Delta_{pg}=350$ cm$^{-1}$ fixed from
our previous analysis\cite{hwang06} and assumed temperature
independent. The length of the remaining ungaped arc is modelled
through the self consistent DOS $\tilde{N}(\omega)$ which we take
to be equal to
$\tilde{N}(0)+(1-\tilde{N}(0))(\omega/\Delta_{ps})^2$ for $\omega
\leq \Delta_{pg}$ and $2(1-\tilde{N}(0))/3$ for $\omega \in
(\Delta_{pg},2\Delta_{pg})$ and 1 beyond this. The value of
$\tilde{N}(0)$ is temperature dependent and assumed to be
proportional to the reduced temperature $t = T/T^{*}$. This agrees
with the idea that the arc closes linearly in reduced temperature
as $T$ is lowered below the pseudogap temperature $T^{*}$. Further
the lost states below $\omega = \Delta_{pg}$ due to the opening of
the pseudogap are taken to reappear above $\Delta_{pg}$ in the
interval up to $2\Delta_{pg}$. The fitting parameters to the data
are area under the resonance peak ($I_p$) and the background
($I_0$).

%
%
\begin{figure}[t]
  \vspace*{-1.5 cm}%
  \centerline{\includegraphics[width=3.5 in]{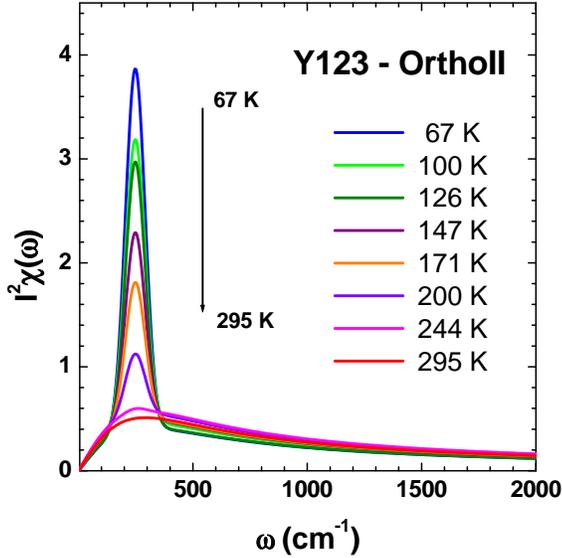}}%
  \vspace*{-2.0 cm}%
\caption{(color online) Model electron-boson spectral density
$I^2\chi(\omega)$ at various temperatures, obtained from our fits
to the scattering rate data assuming an optical resonace peak at
31 meV superimposed on an MMP background.}
  \label{Fig2}
\end{figure}

The fits obtained are seen in the top frame of Fig. \ref{Fig1} and
are seen to be excellent. They are better than the earlier fits
obtained in ref. [20] where a pseudogap was included but without
arcs and recovery region. These fits are reproduced in the lower
frame and are clearly not as good. In particular, in the region
around $\omega =$ 1000 cm$^{-1}$ the new fits capture the narrow
range of variation with temperature observed in the data. This
effect is a direct consequence of the temperature dependence
assumed for the DOS which itself reflects the temperature
dependence of the Fermi surface arcs. When the arcs become smaller
with reducing normalized temperature $t = T/T^{*}$ more states are
lost in the pseudogap region as a results of a reduction in the
value of $\tilde{N}(\omega)$. But in our model for
$\tilde{N}(\omega)$ which is shown in the inset of the upper frame
of Fig. \ref{Fig1}, it is assumed that the lost states are to be
found in the region between $\Delta_{pg}$ and $2\Delta_{pg}$ so as
to respect conservation of states in the electronic DOS. The
existence of this recovery region directly leads to more
scattering {\it i.e.} an increase in $1/\tau^{op}(\omega)$ in this
energy range than would be the case if no recovery occurred {\it
i.e.} $\tilde{N}(\omega)$ was equal to its background value of 1.
This effect pushes up the $1/\tau^{op}(\omega)$ curves more at low
temperatures than at high temperature where the amount of extra
states due to recovery is reduced and finally vanishes in our
model at the pseudogap temperature $T = T^{*}$. It is therefore
through the effect on $1/\tau^{op}(T,\omega)$ of the recovery
region that we see in optics the arcs and their temperature
dependence. The arcs have their greatest effects at low
temperature as can be seen in the middle frame of Fig. \ref{Fig1}
where we show the contribution to the optical scattering rate due
to the recovery states in $\tilde{N}(\omega)$ above $\Delta_{pg}$.
At high temperatures this contribution vanishes as
$\tilde{N}(\omega)$ becomes equal to 1 everywhere. We note that,
with increasing $T$ away from $T_c$, the effect of the recovery
region in $1/\tau^{op}(T,\omega)$ becomes small and we cease to be
able to extract from the data any quantitative information on the
arcs and on the recovery region above it.

In Fig. \ref{Fig2} we show the results of our fits for the
electron-boson spectral function $I^2\chi(\omega)$ as a function
of frequency and temperature. At each temperature the background
was allowed to vary as was the area under the resonance peak fixed
at an energy of 31 meV\cite{stock05}. We see that the peak
disappears with increasing temperature and at 250 K and above only
the background remains. Relative to the peak, the temperature
dependence of the background is small.
%
%
\begin{figure}[t]
  \vspace*{-1.8 cm}
  \centerline{\includegraphics[width=3.5 in]{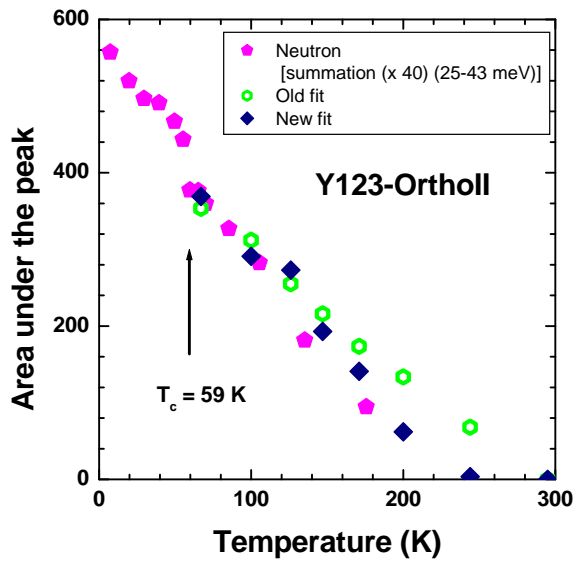}}%
  \vspace*{-2.2 cm}%
\caption{(color online) Comparison of the area under the magnetic
resonance mode seen in inelastic neutron scattering experiments
(solid purple pentagon)\cite{stock05} and the area under the
resonance peak at 31 meV of Fig. \ref{Fig2} (sold blue diamond).
Also shown are the previous fits of ref. [20] (empty green
hexagon) which did not include a recovery region in the electronic
density of state model.}
  \label{Fig3}
\end{figure}

In Fig. \ref{Fig3} we show results for the area under the peak as
a function of temperature above the superconducting $T_c =$ 59 K
{\it i.e.} in the normal state. The points are denoted by (blue)
solid diamond, they are also compared with our previous results
from optics (green)open hexagon\cite{hwang06} and with neutron
results (purple) solid pentagon\cite{stock05}. It is seen that the
fit to the neutron data has improved as compared to our previous
optically derived estimates.

Data on the optical scattering rate in orthoII YBCO$_{6.50}$ show
clear evidence for the formation of a pseudogap which opens at
$T^{*}$ (the pseudogap temperature). We have used a model of the
pseudogap with Fermi arcs with linear variation in reduced
temperature $t = T/T^{*}$ as suggested in ARPES experiments on
underdoped Bi-2212. The fraction of the Fermi surface which
remains ungaped provides a finite but reduced value for the self
consistent density of state at $\omega =$0 with
$\tilde{N}(T,\omega=0)$ linear in $t$. A quadratic dependence of
$\tilde{N}(T,\omega)$ in $\omega$ is assumed till $\omega =
\Delta_{pg}$ after which point we have assumed there is a region
of increased DOS which contains all the missing states so that by
$\omega = 2\Delta_{pg}$ states are conserved. This is a reasonable
assumption and certainly supported by our fit for $T$ near $T_c$.
At higher temperatures however, while the data remains consistent
with the assumed model, details such as the extent in energy and
exact amount of recovery could not be quantitatively pinned down.
With this model we calculate the optical scattering time
$\tau^{op}(\omega)$ at several temperatures in the normal state
with electron-boson spectral density consisting of an MMP
background and resonance peak set at 31 meV to agree with the
available inelastic neutron scattering data on the local magnetic
susceptibility\cite{stock05}. Two parameters, the amplitude of the
background and the area under the resonance peak are varied to get
a best least square fit to the data. The quality of the fit
obtained in this way is considerably superior to an earlier fit
also based on a pseudogap model but without recovery and Fermi
surface arcs. We conclude that the data is more consistent with
arcs than without. The effect of the arc shows up clearly in the
region just above the main rise in scattering rate where all the
curve bunch up together. This is properly captured in our fits
while it is missing in the previous analysis. Also our new fits
agree better with the neutron studies as to the temperature
dependence of the area under the resonance peak which is one of
the important parameter varied in our least square fit.

\acknowledgments This work has been supported by the Natural
Science and Engineering Research Council of Canada(NSERC) and the
Canadian Institute for Advanced Research (CIAR).

\end{document}